# Anisometric Dielectric “Snowman-Particle” Provides Improved Control Over Photonic Hook

Yu. E. Geints[1,*], I.V. Minin[2], O.V. Minin[2]

[1]V.E. Zuev Institute of Atmospheric Optics, 1 Acad. Zuev square, Tomsk, 634021, Russia
[2]Tomsk Polytechnic University, Lenina 36, Tomsk, 634050, Russia
*Corresponding author: ygeints@iao.ru

## Abstract

A novel efficient method for controlling photonic hook (PH) with an anisotropic dielectric microparticle dimer in the «snowman-particle» configuration is introduced and numerically justified. In contrast to existing schemes that rely on structured illumination, partial shielding, or complex morphologies, the proposed anisometric geometry utilizes a simple and easily manufacturable form composed of two partially overlapping cylinders. The finite element method was employed to study how the degree of asymmetry, spatial rotation, and optical properties of the upper particle (the “head” of the snowman-particle) affect the bending angle and maximum intensity of PH. For the first time, it has been demonstrated that adjusting these parameters allows for the deliberate switching between single and double PH modes, as well as controlling the direction and curvature of the hook, including the fragmentation of the focal area. This approach offers high efficiency in utilizing incident radiation, simplicity in fabrication, and broad tunability of PH characteristics. The findings pave the way for the development of compact near-field optical devices, such as optical traps, high-resolution microscopes, sensors, and tools for surface microstructuring.

## Introduction

In various specialized fields, particles with anisotropic morphology are needed to create what are known as photonic hooks (PHs). A PH is a near-field localized electromagnetic beam that propagates along a curved trajectory, typically with a radius of curvature smaller than the wavelength [1]. Over the last 20 years, there has been intensive research into the methods of generating such curved beams and their properties. The photonic hook concept has since expanded to numerous wave domains of different natures. Besides optics, acoustic and plasmonic hooks have also been experimentally demonstrated [2].

The primary mechanism for generating PH involves breaking the symmetry of any particle characteristic. This breaking can result from the particle's geometry, material composition, or illumination conditions, which introduce asymmetry into the phase of the scattered field and consequently bend the energy flow path. At the same time, the morphology of the particle scattering properties is largely determined by its shape [3]. Generating PH using mesoscale particles—those with dimensions on the order of the radiation wavelength—has been achieved through four main mechanisms: (a) selecting an asymmetric particle shape, including composite particles [4] and geometrically truncated symmetrical particles; (b) structured irradiation [5]; (c) rotating the particle [6]; (d) modifying the particle surface with metallic or dielectric materials [7, 8] and creating refractive index asymmetry [9]. The latter method includes, in particular, a cylindrical particle based on eccentric ellipsoids [10], a cylindrical microassembly [11], and a semi-cylindrical morphology [12]. Introducing micro-roughness on the outer surface of a spherical particle also leads to the emergence of PH [13].

The interest in structured beams such as photonic hooks stems from their unique properties and wide range of applications, including optical traps, green photonics, microscopy, and surface microstructuring [1, 14]. For instance, the ability to control and shape curved distributions of optical vortices when illuminating Janus spheres (particles with differing optical properties) was shown in [15].

Although there are many different forms and structures that can be used to create PH, each has its own pros and cons. Recently, a unified approach for generating PH was introduced in [16]. This approach treats external shielding, partial surface coating, and truncation of spherical or cylindrical shapes as physically equivalent schemes of asymmetric illumination with aperture constraints. Truncated particles, which feature internal structural asymmetry, exhibit similar PH characteristics but with higher efficiency in utilizing incident radiation. Particles with external shielding of incident radiation block part of the flux and are less energy-efficient. It is worth noting that Janus particles with two stepped gradient halves of the material [17], as well as hemispheres on a substrate made of another material [18], suffer from parasitic reflection of part of the refracted electromagnetic flux due to the Snell's law when light moves from a more optically dense part to a less dense one. In short, various approaches can produce PH with comparable characteristics, but they differ in energy efficiency and ease of manufacturing.

At the same time, the universe is full of spherical surfaces from raindrops to stars [19-21]. The spherical shape is common in nature because gravity and surface tension minimize the surface area. Moreover, spherical surfaces are quite technologically feasible. In this context, the most promising strategy for creating PH seems to be based on a uniform sphere with introduced shape asymmetry, such as asymmetric dimers of dielectric particles in the form of a snowman, which are widespread, for example, in molecular systems [22, 23]. In this paper, we demonstrate the potential for effectively controlling the morphology of PH by changing the degree of asymmetry of a mesoscale combined particle dimer. Special attention is paid to the morphology of asymmetric particles derived from spherical structures as an independent object of study in PH morphology. It is worth noting that anisotropic (snowman-shaped) dielectric particles can be produced using electrophoretic deposition [24], a two-stage semi-periodic emulsion polymerization process [25], or by rapid solvent exchange and controlled co-precipitation [26].

## Snowman-particle model and the numerical simulation technique

The numerical solution to the problem of optical wave scattering by a microparticle dimer with an asymmetric shape (anisometry) was carried out in a two-dimensional (2D) configuration using the finite element method (FEM), implemented in the COMSOL Multiphysics software package. Using 2D geometry instead of the real three-dimensional (3D) space makes it possible to significantly speed up the numerical calculation and reduce the requirements for computational resources. At the same time, as shown in [27], the artificial reduction in dimensionality in the problem of near-field optical focusing using particles of a curved shape (sphere versus cylinder) preserves all the main features of the field intensity distribution inherent in the photonic nanojet (PNJ) effect, which also applies to PH. e.g., subwavelength transverse dimensions, increased extent, and high intensity. The spatial coordinate transformation in the transition from 3D to 2D changes only the scale and amplitude of the PNJ effect. Therefore, in the following, a dimer consisting of two cylinders of infinite length was used as the snowman-particle (SP), as shown in Fig. 1a.

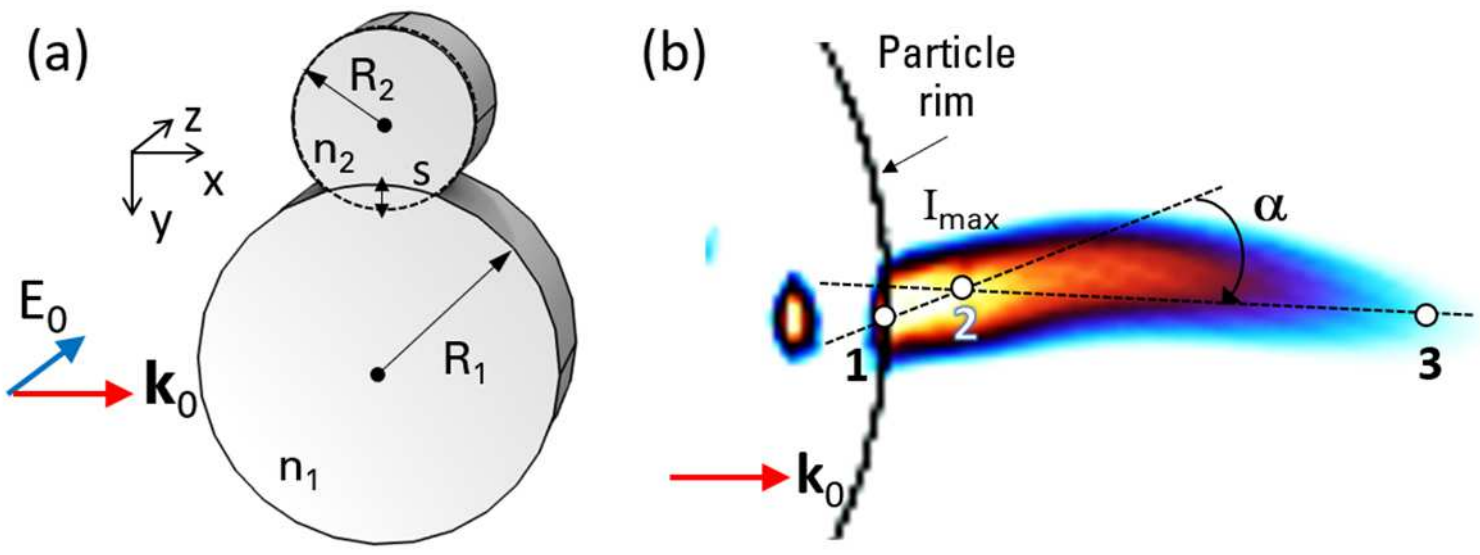

Fig. 1. (a) Geometry of the SP numerical model: a dimer consisting of two infinite cylinders with radii $R_1$ and $R_2$ and refractive indices $n_1$ and $n_2$; $s$ is the mutual overlap parameter. (b) Scheme for determining the bending angle $\alpha$ of the photonic hook: 1 — PH onset point on the particle surface, 2 — intensity maximum $I_{\max}$ position, 3 — the conditional termination of the PH at the $1/e$ level.

A plane monochromatic optical wave with wave vector $\mathbf{k}_0$, propagating along the **x**-axis and having an electric-field amplitude $E_0$ polarized along the **z**-axis, is incident on an asymmetric dimer from the lateral side. In the shadow region of the particle, a curved localized focusing region—a photonic hook—is formed. The PH is characterized by a bending angle $\alpha$ and a maximum relative intensity $I_{\max} = |\boldsymbol{E}_{\max}/E_0|^2$, as shown in Fig. 1b. The angle $\alpha$ is defined as the angle between two straight lines: the first passes through point 1, which is the onset of the PH on the particle surface, and point 2, corresponding to the PH intensity maximum; the second passes through point 2 and point 3, corresponding to the conditional end of the PH, where the intensity decays to $1/e$ of the maximum.

The snowman particle itself is formed by a pair of dielectric cylinders with circular cross sections, having radii $R_1$ and $R_2$ and refractive indices $n_1$ and $n_2$, respectively. In the simulations, partial penetration of one particle into the other was allowed, giving rise to a monolithic conglomerate. The degree of such overlap is governed by a parameter $s$, which specifies the relative arrangement of the cylinders; for $s > 0$, the cylinders partially overlap, forming an asymmetric dimer.

The vector Helmholtz equation for the electric field $\mathbf{E}(\mathbf{r})$ was solved numerically:

$$\nabla\times\nabla\times\mathbf{E}(\mathbf{r}) - k^2\varepsilon(\mathbf{r})\mathbf{E}(\mathbf{r}) = 0 \quad (1)$$

where $\varepsilon(\mathbf{r}) = n^2(\mathbf{r})$ is the coordinate-dependent dielectric permittivity of the medium, $k = 2\pi/\lambda$ is the wave number in the medium, and $\lambda$ is the wavelength of the radiation. In the chosen 2D formulation with field polarization along the **z**-axis, the equation (1) is reduced to a scalar form for the $E_z$ component (TE wave polarization is adopted):

$$\nabla\cdot(\nabla E_z) + k^2\varepsilon(\mathbf{r})E_z(\mathbf{r}) = 0 \quad (2)$$

The optical radiation is assumed to be monochromatic with a wavelength of $\lambda$=532 nm. The SP was placed in air with a refractive index of $n$=1. Absorption of radiation in all materials and media was neglected, i.e., the imaginary part of the refractive index was set to zero. The computational domain was bounded by perfectly matched layers (PMLs) preventing spurious reflection of optical waves from the outer faces of the domain. For discretization of the problem, an adaptive triangular mesh with maximum and minimum edge lengths of 60 nm and 1 nm, respectively, was used, which provided sufficient resolution of the interfaces between media and regions of strong field gradient, including the PH formation region.

## Discussion

Figure 2 presents the two-dimensional distribution of the relative optical field intensity $I = |\boldsymbol{E}/E_0|^2$ and a vector map of the time-averaged energy flux, defined by the Poynting vector $\mathbf{S} = 1/2\,\mathrm{Re}\left[\mathbf{E}\times\mathbf{H}^*\right]$, in the vicinity of an asymmetric SP particle with $R_1$ = 1 µm, $R_2$ = 0.5 µm, and $n_1 = n_2 = 1.5$. On the illuminated side of the particle (left), an interference pattern arising from the superposition of the incident and scattered waves is clearly visible, along with periodic oscillations of the Poynting vector. The key feature here is the formation of a PH in the shadow region to the right of the large particle. The PH manifests itself as an elongated curved region of enhanced intensity with a maximum of $I_{max} \approx 6$. The onset of the PH (point 1) is located on the surface of the large particle near the position with coordinates $x = 0.8$ µm**,** $y = 0$; the PH maximum (point 2) is at $x \approx 1.2$ µm**,** $y \approx 0$, and the conditional endpoint (point 3) lies at a distance of about 1.5 µm from the surface, at $x = 2.5$. The hook axis is bent downward, i.e., in the direction opposite to the upper small particle, which is due to the asymmetry of the dimer. The bending angle α is determined from the three indicated points, as described in the text.

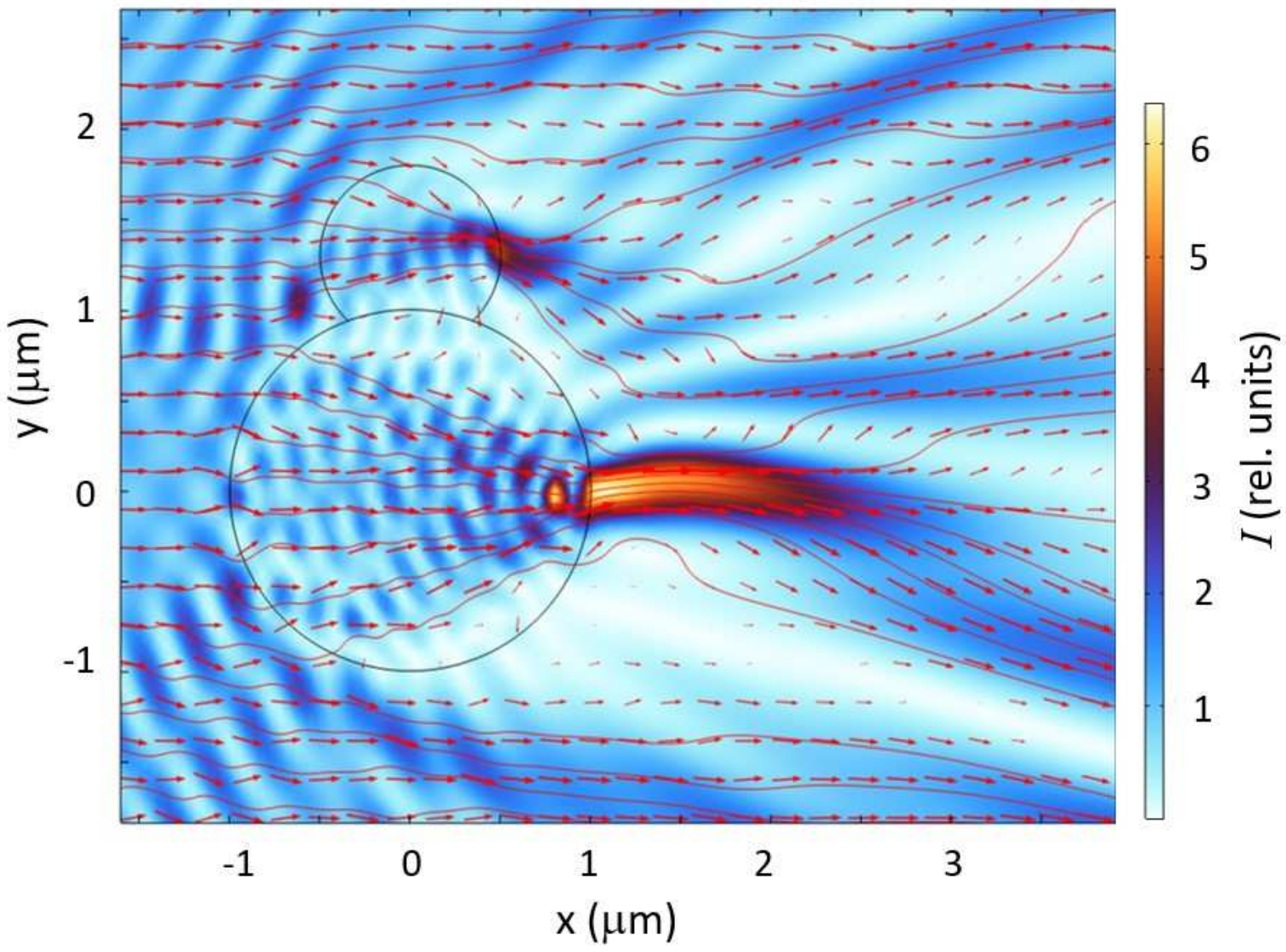

Fig. 2. Two-dimensional distribution of the relative intensity of the optical field $I$ (color scale) and a vector map of the time-averaged energy flux (Poynting vector **S**, red arrows) in the vicinity of the asymmetric SP with $n_1 = n_2 = 1.5$. A plane monochromatic wave propagates from left to right along the x axis (TE polarization). Thin black circles indicate the initial positions of the dielectric cylinders forming the dimer ($R_1$ = 1 µm, $R_2$ = 0.5 µm, $s = 0.2R_2$). In the shadow region to the right of the large particle, a curved PH is formed with a maximum relative intensity $I_{max} \cong 6$. The orientation of the Poynting vector along the curved PH axis indicates a directed transfer of optical energy in this region.

The vector map of energy fluxes qualitatively confirms the directional character of focusing in the PH region. Here, the Poynting vector is oriented predominantly along the curved hook axis, i.e., energy is transferred along the hook rather than merely being localized in a standing wave. Near the surface of the particles, the energy flux flows around the obstacle, being redistributed into the shadow zones and lateral regions. In addition to the main PH, a local intensity maximum is observed at the right edge of the small particle at $x = 0.5$ µm, $y = 1.2$ µm, corresponding to the secondary additional PH. Thus, it is evident that the asymmetric SP forms a curved near-field focus with subwavelength transverse dimensions and a field enhancement of up to six-fold in intensity.

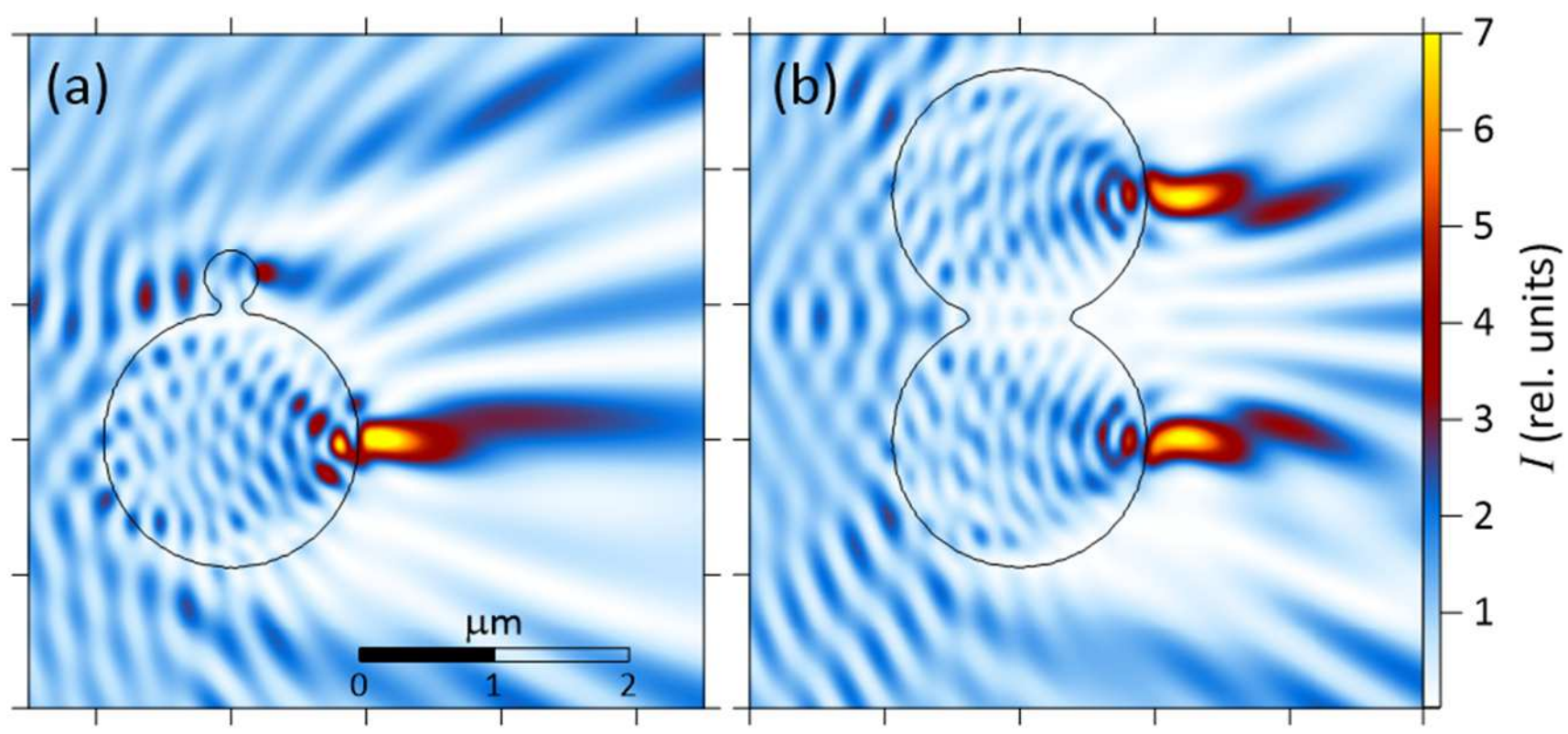

Fig. 3. Effect of the asymmetry of the SP dimer on the spatial shape of the photonic hook. Two-dimensional distributions of the relative optical intensity $I$ in the vicinity of the dimer at $R_1$ = 1 μm ($n_1 = n_2$ = 1.5, λ = 532 nm) for (a) $R_2$ = 0.25 μm representing a strongly asymmetric dimer, and (b) $R_2$ = 1 μm—a symmetric dimer. The incident wave propagates from left to right. In (a), the small upper particle produces only a local field enhancement at its shadow surface, whereas the main extended PH is formed by the lower particle. In (b), both particles give rise to PHs of comparable intensity, forming a *twin* photonic hook.

It should be noted that the emergence of the so-called *twin* PH in dielectric microparticles is associated with a redistribution of optical fluxes either due to the complex geometric shape of the scatterer itself, which creates several regions of field focusing [4], or due to illumination of the particle by structured optical radiation, for example, by two crossed beams [5]. In the situation shown in Fig. 2, the intensity in the secondary (upper) PH is approximately half that in the main focus. Obviously, this is due to the different sizes of the microparticles comprising the dimer.

Indeed, as shown in Fig. 3, the spatial structure of the near-field scattering depends substantially on the degree of asymmetry of the SP dimer, i.e., on the ratio of the radii of the upper and lower particles. In all calculations, the lower particle has a radius $R_1$ = 1 μm, and the refractive indices of both particles are identical ($n_1 = n_2$ = 1.5). At $R_2$ = 0.25 μm (Fig. 3a), the upper particle, owing to its subwavelength size, proves incapable of forming an extended directional photonic hook, and only a local intensity maximum with $I_{max} \approx 4$ arises in its shadow region, which decays rapidly and does not form an elongated focal region. The main PH is formed by the lower, larger particle and has a maximum relative intensity of $I_{max} \approx 7$. Its extent is approximately from 2 to 2.5 μm, and its shape exhibits a slight downward tilt without bending of the photonic flux. The secondary (upper) maximum turns out to be approximately half as strong as the main one, which is qualitatively consistent with Fig. 2. As a result, under strong asymmetry of the SP dimer, the upper particle plays the role of a weak local scatterer that merely perturbs the field of the main focal region but does not create a full-fledged second PH.

When the radius of the upper particle is increased to $R_2$ = 1 μm (Fig. 3b), the dimer becomes symmetric, i.e., both particles have identical radii and refractive indices. In this case, each of them forms its own photonic hook of comparable intensity ($I_{max} \approx 7$). In the shadow region, a *twin* photonic hook is observed in the form of two pronounced focal regions propagating along the **x**-axis and separated along the **y**-coordinate. The interaction of the two PHs, arising from the superposition of the waves scattered by the upper and lower particles, gives rise to interference modulation of the optical field. Thus, the relative intensities of the PHs are governed by the asymmetry of the SP dimer, when under strong asymmetry, the secondary PH is suppressed, whereas for a symmetric dimer, a twin PH arises with two focal regions of comparable intensity.

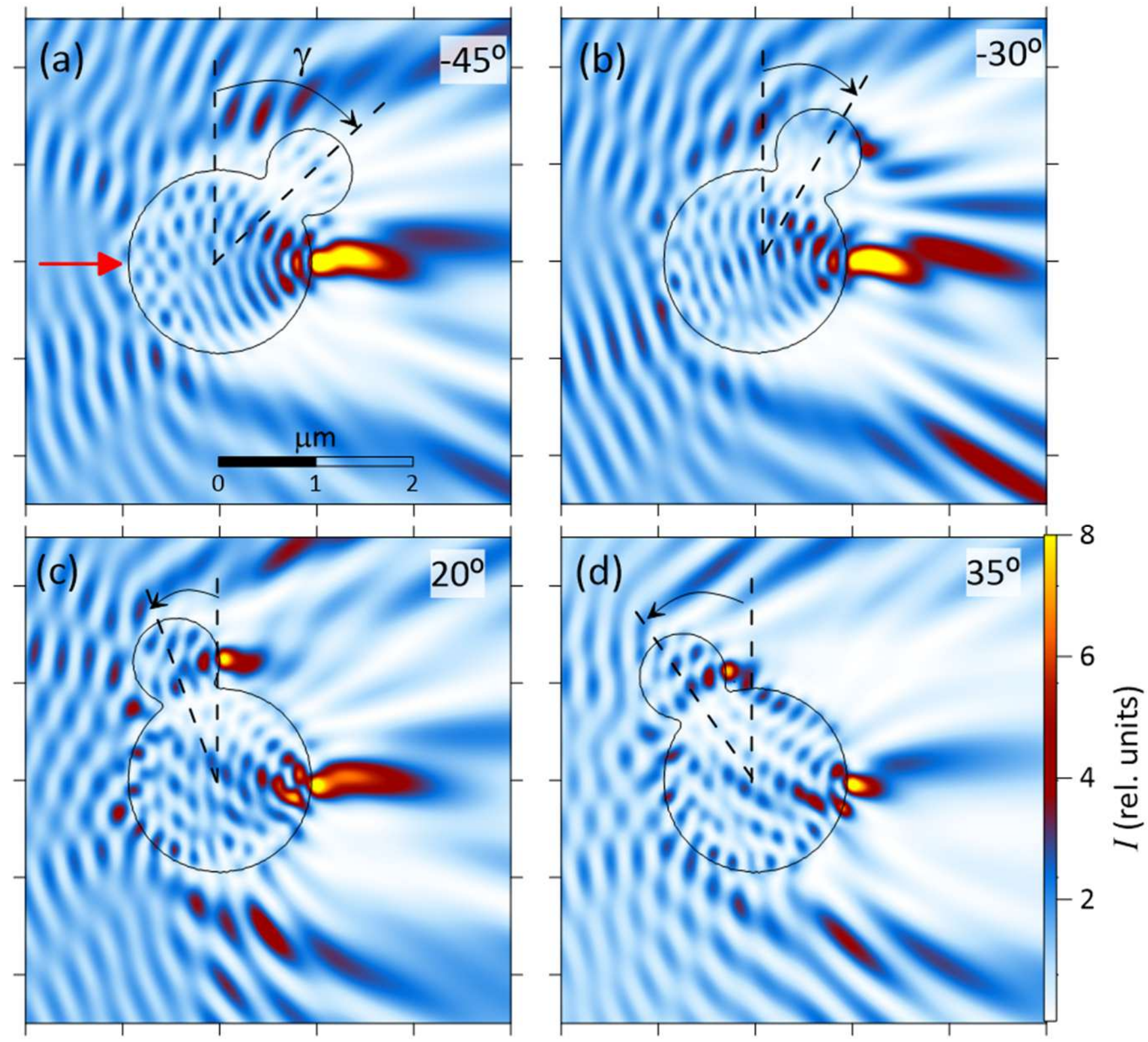

Fig. 4. The effect of spatial rotation (γ) of the SP dimer on the shape of the photonic hook. Two-dimensional distributions of the relative intensity of the optical field $I$ in the vicinity of the dimer with $R_1$ = 1 µm and $R_2$ = 0.5 µm at (a) γ = -45°, (b) γ = -30°, (c) γ = 20°, (d) γ = 35°.

Fig. 4 shows how the spatial orientation of an asymmetric SP dimer relative to the direction of the incident wave governs the shape, direction, and intensity of the PH. A dimer formed by a large lower cylinder with $R_1$ = 1 µm and a small upper cylinder with $R_2$ = 0.5 µm, having identical refractive indices, is considered. Positive values of the rotation angle γ correspond to a tilt of the small particle to the left (counterclockwise), while negative values correspond to a tilt to the right (clockwise).

In panel (a), at γ = -45°, the dimer is strongly tilted to the right. Here, the main PH is formed by the lower larger particle, has the highest intensity ($I_{max}$ ≈ 8), is elongated to the right almost horizontally, and is only weakly curved. The contribution of the upper small particle manifests itself as a local perturbation of the field near its shadow surface but does not create a pronounced second hook. As the negative rotation angle decreases to γ = -30°, shown in panel (b), the main PH retains its high intensity and extent; however, its axis and curvature change somewhat, and the field intensity near the small particle becomes higher, which is associated with a change in the interference conditions of the scattered waves.

On shifting to positive angles, the field distribution is substantially rearranged. In Fig. 4(c), at γ = 20°, the small particle is displaced to the left, and the main PH formed by the lower particle acquires an upward tilt. At the same time, its intensity decreases somewhat, and the spatial distribution acquires a pronounced curvature. Finally, in panel (d), at γ = 35°, the main PH is noticeably weakened and fragmented, which is manifested in the fact that, instead of a single extended bright region, several local maxima are observed, located both at the surface of the large particle and near the small one. This indicates that, under a strong negative rotation of the dimer, the condition for efficient near-field focusing is violated, and the energy is redistributed among several competing regions.

Thus, Fig. 4 clearly demonstrates that the spatial rotation angle γ of the SP dimer acts as an effective control parameter for the photonic hook. By varying this angle, one can deliberately change the bending direction of the PH, its length, and its maximum intensity, and at large positive values of γ, one can achieve splitting or even fragmentation of the focal region. This opens up additional possibilities for controlling near-field optical fields by means of asymmetric dielectric microparticles.

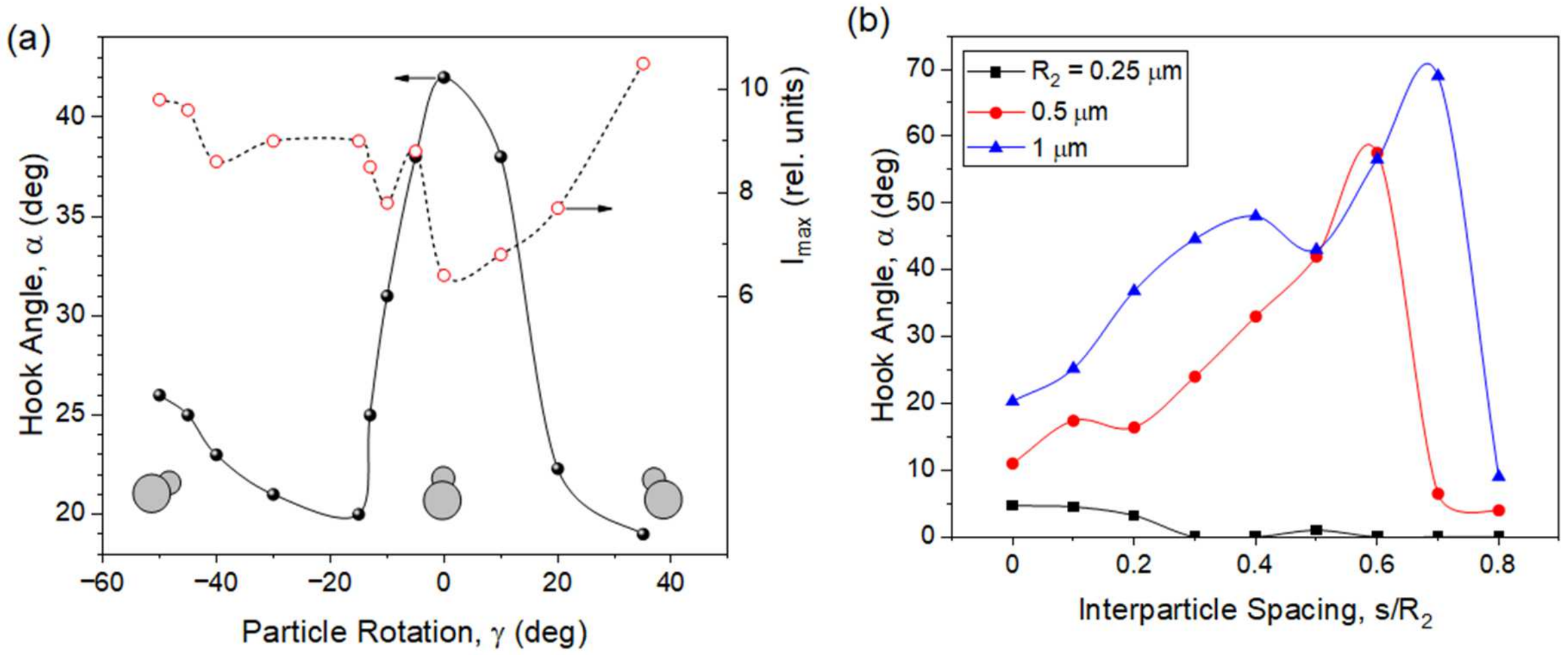

Fig. 5. Dependence of the bending angle of the photonic hook α and its maximum relative intensity $I_{max}$ on the geometric parameters of the SP dimer with $R_1$ =1 μm and $n_1$=$n_2$=1.5. (a) Effect of the spatial rotation angle γ ($R_2$ =0.5 μm); the orientations of the dimer are shown schematically in the lower part of the panel. (b) Dependence of the bending angle α on the dimensionless interparticle distance $s/R_2$ for dimers with different $R_2$ (γ =0).

Fig. 5 presents the quantitative characteristics of the photonic hook formed by the SP dimer. Here, the bending angle α and the maximum relative intensity $I_{max}$ are plotted. As can be seen, the dependence α(γ) in panel (a) is nonmonotonic, with a global maximum of α =42–43° in the vicinity of dimer rotation angles γ = -5 to 10°. Away from this interval, the value of α decreases, reaching a local minimum of α ≈ 20° at γ = −15°, after which a repeated increase is observed. The dependence $I_{max}$(γ) is also nonmonotonic, with the minimum intensity values $I_{max}$ = 6 corresponding to the region of maximum PH bending at γ = 0, whereas at values |γ| ≳ 30–40°, the intensity smoothly increases to a value of ~10 rel. units. Thus, the conditions for achieving maximum PH curvature and maximum field intensity do not coincide, which indicates the existence of a trade-off between the geometric and energy characteristics of the formed hook.

Fig. 5b shows the effect of the dimensionless interparticle distance $s/R_2$ on the PH bending angle α for three values of the upper particle radius $R_2$. At $R_2$ = 0.25 μm, the angle α does not exceed 5° over the entire considered range of the parameter $s$, which indicates the inability of such a small particle to form a pronounced curved PNJ. For a larger upper particle of the dimer with $R_2$ =0.5 μm, the dependence α($s$) under consideration is characterized by a smooth increase with a maximum of α ≈ 58° at $s/R_2$ = 0.6 and a subsequent sharp decrease. For $R_2$ = 1 μm, when the SP dimer becomes symmetric, the maximum PH bending angle reaches α ≈ 70°, and the optimal value of the parameter $s/R_2$ shifts to ~0.7. In all cases, after the maximum, a rapid decrease in α is observed, caused by the weakening of the interparticle interaction and the violation of the conditions for the formation of a single curved focal region.

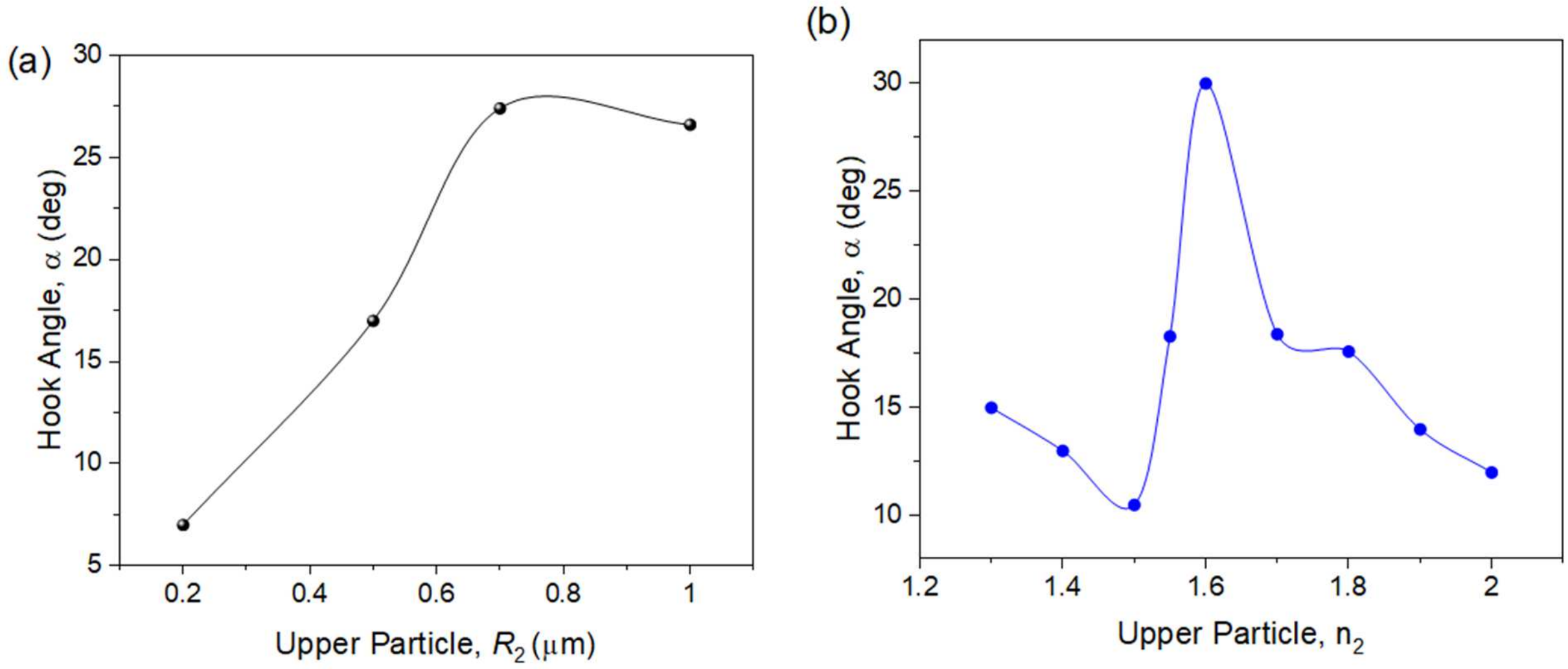

Fig. 6. Dependence of the bending angle of the photonic hook α on the parameters of the upper particle of the SP dimer ($R_1$ = 1 μm, $n_1$ =1.5): (a) on the radius $R_2$ at a fixed refractive index $n_2$=1.5; (b) on the refractive index $n_2$ at a fixed radius $R_2$ = 0.5 μm.

The simulation results presented in Fig. 3 above show that the radius of the upper particle of the SP dimer is one of the effective control parameters of the PH. Varying $R_2$ provides a smooth tuning of the bending angle with saturation at large sizes. Fig. 6a presents quantitative data on the effect of the size of the upper particle of the SP dimer on the bending angle α of the formed PH. In all calculations, the lower particle of the dimer retained fixed parameters: $R_1$ = 1 μm, $n_1$ = 1.5. It can be seen that, as $R_2$ increases from 0.2 to 0.7 μm, the PH bending angle monotonically increases from α ≈ 6.5° to ~27.5°. A further increase in the radius of the upper particle to 1 μm, with the formation of a symmetric SP dimer, leads to an insignificant decrease in the hook bending angle α to a value of ~26.5°, which indicates that the dependence reaches saturation. Thus, it can be noted that, in the range of upper-particle radii around $R_2$ ≈0.7 μm, the maximum curvature of the forming PH is realized. At smaller radii, the upper particle does not create a sufficient perturbation of the field, whereas at larger ones, the effect stabilizes.

Fig. 6b shows the dependence of α on the refractive index of the upper particle $n_2$ at a fixed radius $R_2$ =0.5 μm. As can be seen, this dependence has a clearly pronounced non-monotonic character. As $n_2$ increases from 1.3 to 1.5, the PH bending angle α first decreases from ~15° to a minimum value of ~10.5°, after which it sharply increases, reaching a maximum of α ≈ 30° at $n_2$ = 1.6. A further increase in $n_2$ to 2.0 is accompanied by a rapid decrease in the angle α to 12°. The presence of a quasiresonant maximum near the value $n_2$ =1.6 indicates a significant role of interference and diffraction effects in the internal field of the upper particle, since it is precisely in this range of refractive index values that the most favorable conditions are created for the redistribution and interaction of optical fluxes from different parts of the dimer, with the formation of a pronounced curved PH.

## Conclusion

The numerical simulation performed convincingly demonstrates that an asymmetric dielectric snowman-type dimer particle represents an efficient and technologically feasible tool for controlling the photonic hook which, unlike many known approaches, is based on a simple geometric shape in the form of two partially overlapping cylinders, thereby ensuring high energy efficiency and ease of fabrication. Varying the degree of asymmetry, the spatial orientation, and the optical characteristics of the upper particle of the dimer allows the bending angle, the extent, and the intensity of the formed PH to be deliberately tuned. In particular, under strong asymmetry, the secondary hook is suppressed, whereas for a symmetric dimer, a *twin* PH arises with two near-field focal regions of comparable intensity. Rotation of the dimer about the center of the main particle provides smooth tuning of the bending direction and, at large angles, leads to splitting or fragmentation of the focal region. The established dependences of the PH bending angle on the parameters of the upper particle are nonmonotonic, and the maximum curvature and the maximum PH intensity are achieved at different values of the control parameters. This is important to take into account when optimizing near-field optical devices. We note that further optimization of the photonic hook characteristics and, consequently, of the particle structure can be possible using various artificial intelligence (AI) methods [28].

The body of results obtained demonstrates that the proposed approach based on asymmetric dielectric dimers opens up additional possibilities for controlling subwavelength structured optical fields and may be in demand in problems of optical micromanipulation, near-field microscopy, and surface structuring; it also expands the range of asymmetric mesoscale particles available for the formation of structured localized fields. The results of this work open new horizons for the creation of compact near-field optics devices, and the demonstrated possibility of controlling the PH morphology through the internal parameters of the particle creates prerequisites for the integration of such dimers into photonic chips and lab-on-a-chip systems, and may lead to the creation of adaptive and tunable photonic elements of a new generation.

**Acknowledgements.** Y.G acknowledges the support from Ministry of Science and Higher Education of the Russian Federation. I.M. and O.M. are partially supported in the framework of the Tomsk Polytechnic University Competitiveness Enhancement Program, Russia.

**Funding.** Ministry of Science and Higher Education of the Russian Federation (IAO SB RAS).

**Disclosures.** The author declares no conflicts of interest.

**Data availability.** Data underlying the results presented in this paper may be obtained from the authors upon reasonable request.